\documentclass[letterpaper,twocolumn,prl,aps,amsmath,amssymb,floatfix,longbibliography]{revtex4-1}
\usepackage{mathptmx}
\usepackage{color}
\usepackage{amsmath}
\usepackage{amssymb}
\usepackage{graphicx}
\usepackage{esint}
\usepackage[normalem]{ulem} 
\usepackage[unicode=true, bookmarks=true,bookmarksnumbered=false,bookmarksopen=false,
 breaklinks=false,pdfborder={0 0 1},backref=false,colorlinks=true]{hyperref}
\hypersetup{linkcolor=magenta,urlcolor=blue,citecolor=blue,pdfstartview={FitH},hyperfootnotes=false}

\makeatletter

\usepackage{textcomp}
\usepackage{epstopdf}

\pdfpageheight\paperheight
\pdfpagewidth\paperwidth

\@ifundefined{textcolor}{}{%
 \definecolor{BLACK}{gray}{0}
 \definecolor{WHITE}{gray}{1}
 \definecolor{RED}{rgb}{1,0,0}
 \definecolor{GREEN}{rgb}{0,1,0}
 \definecolor{BLUE}{rgb}{0,0,1}
 \definecolor{CYAN}{cmyk}{1,0,0,0}
 \definecolor{MAGENTA}{cmyk}{0,1,0,0}
 \definecolor{YELLOW}{cmyk}{0,0,1,0}
}

\usepackage{xcolor}
\usepackage{soul}
\newcommand{\ket}[1]{\ensuremath{\left|#1\right\rangle}}

\definecolor{blue}{rgb}{0,0,1}
\definecolor{red}{rgb}{1,0,0}
\definecolor{green}{rgb}{0,1,0}

\usepackage{soul}

\makeatother

\begin{document}

\title{Coherent optical control of a thousand qubits in a three-dimensional trapped-ion crystal}

\author{Yue Li$^{1}$}
\thanks{These authors contributed equally to this work.}
\author{Yuxiang Huang$^{1,2}$}
\thanks{These authors contributed equally to this work.}
\author{Yi Li$^{1,2,3}$}
\author{Xu Cheng$^{1,2}$}
\author{Zihan Xie$^{1,2}$}
\author{Qingyuan Mei$^{1}$}
\author{Chang-Ling Zou$^{2,4,5}$}
\author{Ya Wang$^{1,2,6}$}
\author{Xinhua Peng$^{1,2,6}$}
\author{Yiheng Lin$^{1,2,6}$} \email{yiheng@ustc.edu.cn}
\author{Jiangfeng Du$^{7}$}

\maketitle

\onecolumngrid
\begin{center}
\small
\begin{minipage}{0.88\textwidth}
\centering
\vspace{-3.5em}
\noindent
\textit{
$^1$Laboratory of Spin Magnetic Resonance, School of Physical Sciences, Anhui Provincial Key Laboratory of Scientific Instrument Development and Application, University of Science and Technology of China, Hefei 230026, China\\
$^2$Hefei National Laboratory, University of Science and Technology of China, Hefei 230088, China\\
$^3$National Advanced Talent Cultivation Center for Physics, University of Science and Technology of China, Hefei 230026, China\\
$^4$Laboratory of Quantum Information, University of Science and Technology of China, Hefei 230026, China\\
$^5$Anhui Province Key Laboratory of Quantum Network, University of Science and Technology of China, Hefei 230026, China\\
$^6$Hefei National Research Center for Physical Sciences at the Microscale, University of Science and
Technology of China, Hefei 230026, China\\
$^7$State Key Laboratory of Ocean Sensing and School of Physics, Zhejiang University, Hangzhou 310058, China
}
\end{minipage}
\end{center}

\begin{center}
\begin{minipage}{0.8\textwidth}
\noindent

Leading quantum platforms are largely limited to one or two dimensions, whereas three-dimensional Coulomb crystals of trapped ions self-assemble into dense qubit arrays with no rearrangement overhead. Coherent optical control of such crystals has nonetheless been out of reach, because the optical transition couples to nearly all $3N$ collective modes, i.e., a thermally occupied quasi-continuum bath. Here we cool about 3,600 modes of a crystal of over 1,200 ions below single-phonon occupation, using a continuous cooling band that spans more than five octaves in  vibrational spectrum, from about 50 kHz to 2.1 MHz. The crystal then serves as a register of optical qubits, allowing global initialization, coherent manipulation and collective detection. We demonstrate ensemble optical coherence of 21.6 milliseconds and reduce the optical-transition frequency uncertainty from 389 Hz for a single ion to 12 Hz for over a thousand ions, approaching the expected projection-noise scaling. Our approach could be extended to even larger crystals with phonon-mediated many-body couplings, opening a route towards entanglement-enhanced precision metrology and three-dimensional quantum simulation, particularly with spatial-shaped illumination for region-resolved collective control.

\end{minipage}
\end{center}

\twocolumngrid

\maketitle
\noindent \textbf{\large{}Introduction}{\large\par}
\noindent 

Massive arrays of qubits are a shared goal of every quantum platform, and the leading architectures have so far developed in one and two dimensions~\cite{GoogleQAI2025, Bluvstein2025, Bluvstein2024, Ransford2026, Guo2024}. The number of controllable qubits in trapped-ion systems has grown rapidly along this line over the past two decades. One-dimensional linear chains established quantum simulation and coherent control at the scale of tens of ions~\cite{kim2010,islam2013,Monroe2021}, and two-dimensional crystals have extended these capabilities to hundreds, from collective spin dynamics with about two hundred ions in a Penning trap~\cite{Bohnet2016} to site-resolved quantum simulation with up to 512 ions in monolithic Paul traps~\cite{Kiesenhofer2023,Guo2024,Guo2025}, supported by sub-Doppler cooling methods tailored to their structured motional spectra~\cite{Lechner2016,Jordan2019,Chen2020,Feng2020,Joshi2020,Qiao2021}. Three-dimensional (3D) Coulomb crystals self-assemble into dense, defect-free arrays under the ions' mutual repulsive forces~\cite{Hornekaer2002,Dubin1999,Thompson2015,Morigi2025}, and have been applied in microwave clocks with up to $10^{5}$ ions~\cite{Zuo2019,Burt2021}. 

However, 3D trapped-ion systems have remained beyond the reach of the coherent optical control that is routine in lower dimensions~\cite{Morigi2025}. A crystal of  $N$ ions carries $3N$ collective vibrational modes, which for $N>1,000$ form a quasi-continuum frequency spectrum with a sub-kilohertz mean spacing. As a result, individual modes cannot be resolved, precluding resolved-sideband cooling~\cite{Leibfried2003}. Conventional laser cooling of large 3D crystals has consequently stalled at the Doppler limit, where thermal vibrational excitation inhomogeneously broadens the optical transition and makes high-fidelity manipulation of optical clock qubits increasingly difficult as the crystal size grows. Overcoming this challenge would make it feasible to employ large 3D crystals in scaled quantum computing~\cite{Wu2021}, many-body quantum simulation~\cite{Guo2024}, advanced optical clocks~\cite{Leibrandt2024} and precision spectroscopy with sympathetically cooled molecular ions~\cite{Alighanbari2020}.

Here we demonstrate coherent optical control of a 3D trapped-ion crystal containing $N_{\mathrm{ion}}=1{,}265\pm35$ ions in a
cryogenic radio-frequency Paul trap. We overcome the cooling obstacle by synthesizing a continuous cooling band from electromagnetically induced transparency with parallel resonances~\cite{Morigi2000,Zhang2022}, and sideband spectroscopy~\cite{Leibfried2003,Vybornyi2023} confirms samples of 205  mode-frequency probes in a quasi-continuous spectrum from about 3,600 modes are cooled below single-phonon occupation. The crystal then serves as a register of optical clock qubits with global initialization, coherent manipulation and collective readout, showing high-contrast Rabi oscillations and a coherence time of up to 21.6~ms. The statistical uncertainty of the carrier-transition frequency is reduced from 389~Hz for a single ion to 12~Hz for the full crystal, following projection-noise scaling across three decades in ion number. This apparatus could be extended to far larger crystals, and spatially shaped illumination would add region-resolved collective control with phonon-mediated couplings, opening a route towards entanglement-enhanced metrology and 3D quantum simulation.

\begin{figure*}[tb] 
    \centering
    \includegraphics[width=1.8\columnwidth]{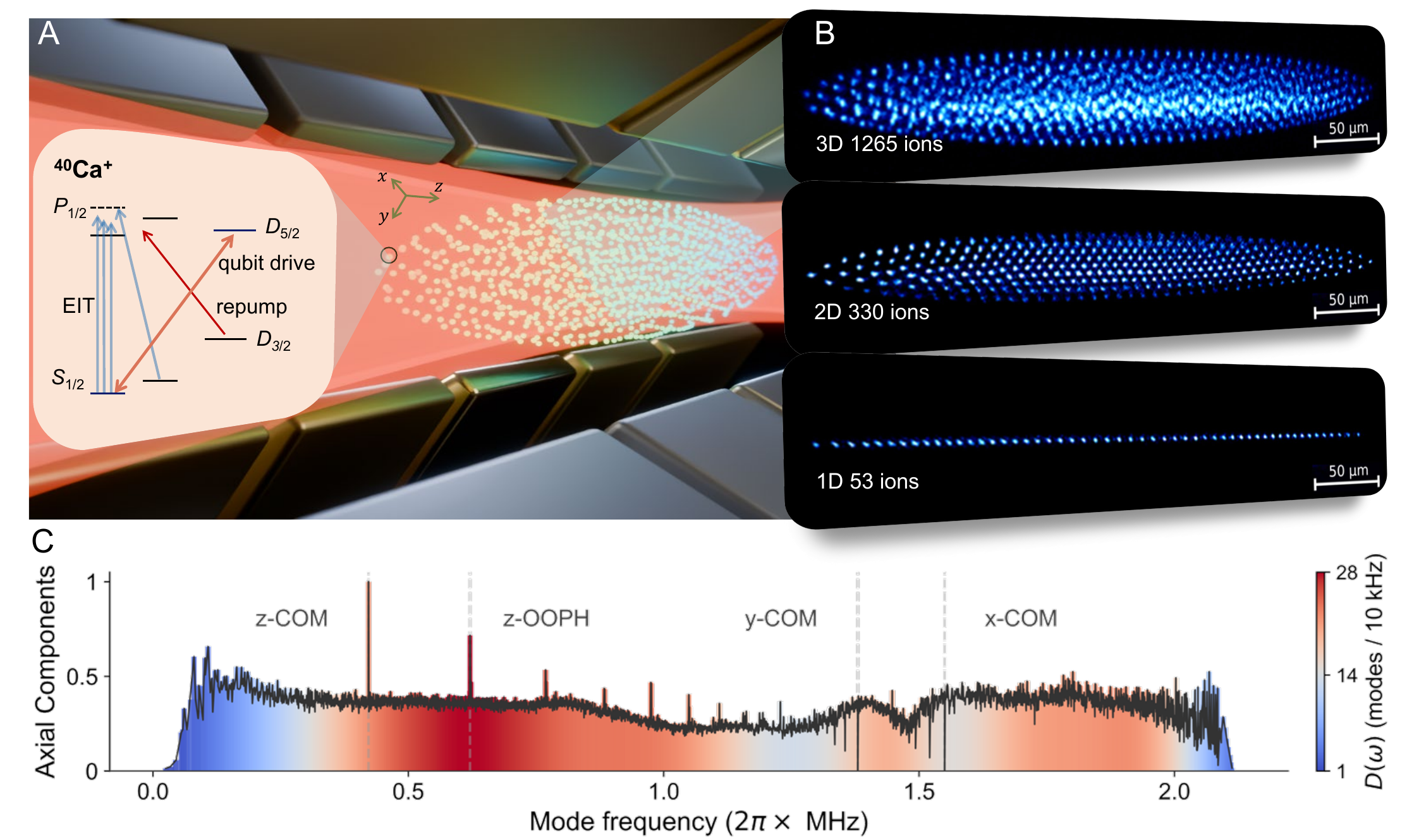}
    \caption{\textbf{Experimental platform and collective vibrational spectrum of a large three-dimensional Coulomb crystal.}
  \textbf{a}, Schematic of the cryogenic Paul trap and optical drives. We utilize multiple 397~nm lasers driving the $S_{1/2}$ to $P_{1/2}$ transition, forming parallel resonances for the EIT cooling. A coherent drive laser of 729~nm is applied for either sideband thermometry, coherent Rabi oscillation or Ramsey coherence interferometry, depending on the laser frequency detuning from resonance. \textbf{b}, Fluorescence imaging along the transverse direction gives the ion crystal configuration. Varied confinements along the three directions give linear, planar and cigar-shaped 3D ion crystal configurations, respectively, each with an axial length of approximately 280 $\mu$m. 
  \textbf{c}, Axial projection components of the collective vibrational modes obtained from numerical simulations, with trap potentials as of 3D crystal in (b). The axial laser couples to the majority of the three-dimensional normal modes, including both axial and radial excitations, except for the center-of-mass modes along radial directions, denoted as x- and y-COM. High projections are observed for the global axial modes, denoted as z-COM and z-OOPH as described in the text. The color shows the mode density $D(\omega)$, revealing the dense quasi-continuous vibrational spectrum of the crystal.
  }
  \label{fig1}
\end{figure*} 

\begin{figure*}[tb] 
    \centering
    \includegraphics[width=2.0\columnwidth]{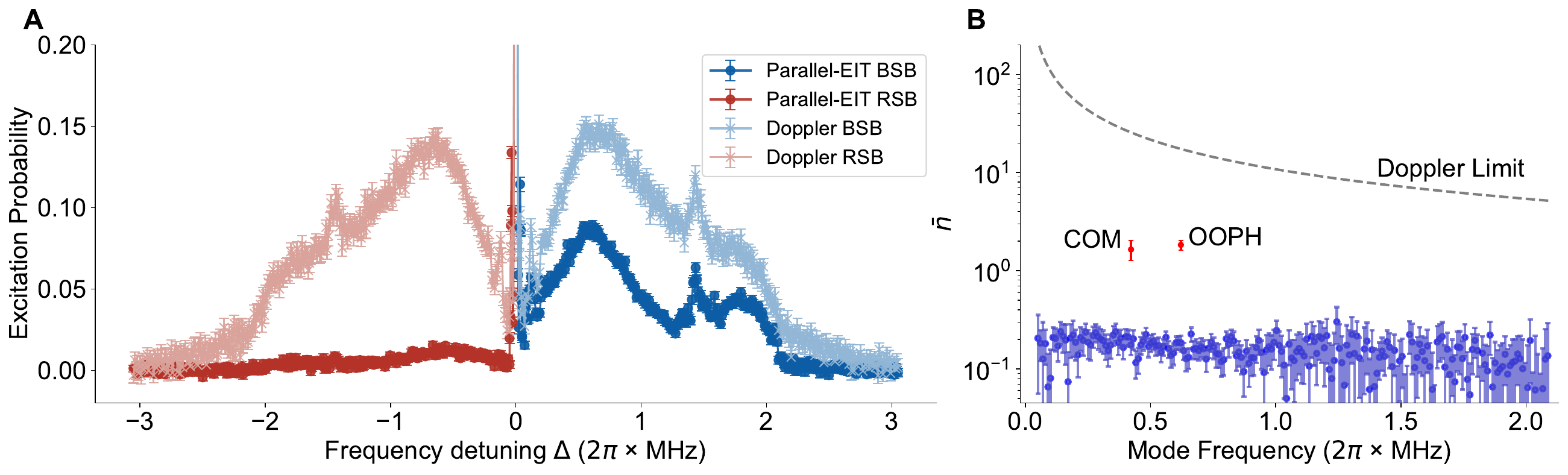}
    \caption{\textbf{Broadband near-ground-state cooling of a dense three-dimensional vibrational spectrum.}
    \textbf{a}, Red-sideband and blue-sideband excitation spectra after Doppler cooling and parallel-EIT cooling. Parallel-EIT produces enhanced sideband asymmetry over the entire measured frequency range, demonstrating broadband suppression of motional excitation. 
    \textbf{b}, Mean phonon number estimated from sideband asymmetry after parallel-EIT cooling $\bar{n} = p_r/(p_b - p_r)$. Swept-frequency ($2\pi\times$10~kHz steps) measurements from $2\pi\times$50~kHz to $2\pi\times$2.1~MHz reveal near-ground-state cooling ($\bar{n} < 0.3$) across most of the vibrational spectrum. The axial center-of-mass and out-of-phase modes are independently characterized due to their narrow linewidths ($\bar{n}=1.7(4)$ and $1.8(2)$).
    }
    \label{fig2}
\end{figure*} 

\noindent \textbf{\large{}Results}{\large\par}

\noindent
\textbf{Large 3D crystal and motional spectrum.}
We prepare a large-scale 3D Coulomb crystal of $^{40}\text{Ca}^+$ ions containing $N_{\rm{ion}} = 1{,}265 \pm 35$ ions in a cryogenic radio-frequency Paul trap held at approximately 4~K temperature (Fig.~\ref{fig1}a). The total ion number and the shape of the crystal are estimated by molecular-dynamics simulation ~\cite{Drewsen1998, sm}, where the best match is found from both the measured axial extent $281(4)$ $\mathrm{\mu}$m (Fig.~\ref{fig1}b) and a harmonic potential that gives single ion secular motional frequencies $\{\omega_x,\omega_y,\omega_z\}/2\pi = \{1.55,1.38,0.422\}\,\mathrm{MHz}$. We find the ions naturally form a fully three-dimensional, cigar-shaped crystal elongated axially with a radial extent of 34.8(3) and 45.1(3) $\mathrm{\mu}$m , with stronger confinement along the transverse direction. The cryogenic ultra-high-vacuum environment ($\sim 10^{-13}$~mbar) provides stable trapping \cite{Spivey2022} with suppressed motional heating, enabling efficient cooling and spectroscopy of the large ensemble.

The internal Coulomb coupling among the ions in the 3D crystal, combined with the trap potential, produces normal motional modes with non-uniform vibration directions across the ions, except for the center-of-mass (COM) modes which vibrate purely along the $x, y$ or $z$ directions. A probe laser along the $z$ direction therefore couples to the majority of the 3D modes. As shown in Fig.~\ref{fig1}c, we characterize the motional spectrum by numerically calculating a 3D crystal of $\Tilde{N}_{\rm{ion}}=1{,}265$ ions in the potential measured above, with $3\Tilde{N}_{\rm{ion}}=3{,}795$ normal modes together with their projections onto the axial direction~\cite{Landa2012}. The resulting mode-density spectrum reveals a densely congested, quasi-continuous manifold with mode frequency $\omega_m/2\pi$ spanning from about $30$~kHz to $2.1$~MHz, with an average mode spacing of approximately $0.5$~kHz. The mode density $D(\omega)$, defined as the number of modes per 10 kHz, ranges from approximately 1 to 28. There are a large number of  equilibrium configurations that may affect the motional modes. We numerically verified across 200 independent equilibrium realizations that the overall axial-projection envelope remains robust, with standard deviations up to 17\% and on average 3.4\%. Sampled over 200 configurations, we obtain a mode density of  1.8(6) per 1~kHz, so the spectrum can be sampled efficiently with a 10~kHz step. We also find macroscopic modes that have negligible frequency deviations despite varied crystal configurations, such as the $z-$direction center-of-mass (COM) and out-of-phase (OOPH) modes (See Supplementary Information~\cite{sm}).

\begin{figure*}[tb] 
    \centering
    \includegraphics[width=1.9\columnwidth]{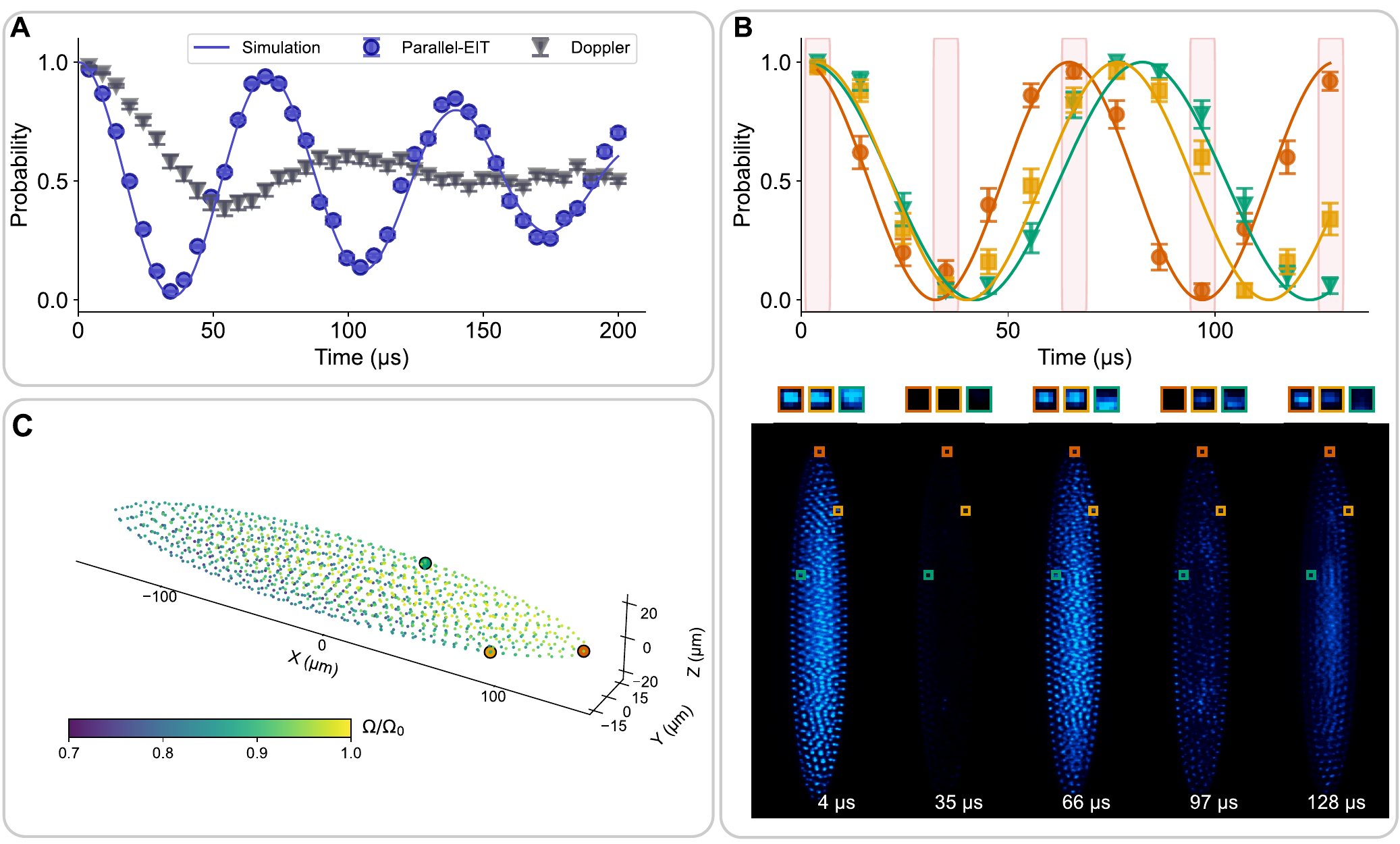}
    \caption{\textbf{Coherent optical control of a three-dimensional ion crystal after cooling.}
    \textbf{a} Global carrier Rabi oscillations measured by collecting fluorescence from the entire ion crystal with a photomultiplier tube (PMT), after conventional Doppler cooling or parallel-EIT cooling. The solid curve represents a simulation for the latter case, based on the above spatial distribution of the Rabi frequency, without incorporating motional-temperature effects or decoherence, and quantitatively reproduces the observed ensemble decay.
    \textbf{b} Rabi oscillation across the ion crystal. We observe high contrast Rabi oscillation within the ion crystal, from fluorescence images that correspond to the spin population at varied evolution duration. Due to the spatially varying intensity profile of the drive laser, we observe varied Rabi rates within the ion crystal, with three spots as examples. Pink bars show the duration stamp corresponding to the above image.
    \textbf{c} Reconstructed Rabi rate distribution across the crystal. The beam waist and the incidence direction are obtained by maximum-likelihood fitting to the Rabi rates of 11 resolved ion spots, and are combined with the reconstructed crystal shape and a Gaussian laser profile.
    }
    \label{fig3}
\end{figure*} 

\noindent
\textbf{Broadband sub-Doppler cooling}
To cool this broad motional spectrum, we first apply 6~ms of Doppler cooling, followed by 10~ms of electromagnetically induced transparency (EIT) cooling with parallel resonances~\cite{Morigi2000,Zhang2022} (detailed in Supplementary Information~\cite{sm}). The parallel resonances arise from a $\pi$-polarized 397 nm probe beam carrying three frequency components, each of which forms a $\Lambda$-type three-level system together with the $\sigma^-$-polarized 397 nm pump beam as shown in Fig.\ref{fig1}a. The two-photon resonance of each system creates a dark state, and the resulting three resonances act as multiple cooling windows distributed across the motional spectrum, providing broadband and sub-Doppler cooling without resolving individual modes. To characterize the motional excitation, we use red- and blue-sideband (RSB and BSB) spectroscopy \cite{Vybornyi2023} on the $\lvert S_{1/2},m=-1/2\rangle\leftrightarrow\lvert D_{5/2},m=-1/2\rangle$ optical transition (Fig.~\ref{fig1}a), driven by the axial 729~nm laser with varied negative and positive frequency detunings with a step-size of 10 kHz for efficient sampling of the spectrum envelope, and its intensity set by a carrier Rabi frequency of  $\Omega/2\pi = 15$~kHz. The pulse duration is adjusted at each spectroscopy frequency such that the sideband excitation probability remains below 0.1 (details in Source Data). Subsequent global fluorescence detection with a photomultiplier tube (PMT) gives the respective RSB and BSB spectra, whose asymmetry reflects motional excitation~\cite{Vybornyi2023}. 

Following parallel-EIT cooling, a pronounced sideband asymmetry develops across the entire probed bandwidth compared with Doppler cooling results (Fig.~\ref{fig2}a), demonstrating broadband suppression of motional excitation. From the measured asymmetry, we estimate the mean phonon excitation $\bar{n}=\frac{p_r}{p_b-p_r}$ where $p_r$ and $p_b$ denote the RSB and BSB excitation probabilities, respectively. 
As shown in Fig.~\ref{fig2}b, we obtain $\bar{n}<0.3$ over most of the sampled frequencies, as validated by excitations in the BSB, indicating broadband near-ground-state cooling across the collective motional manifold, except for the relatively more stable macroscopic $z-$COM and $z-$OOPH modes (as shown in Fig.~\ref{fig1}). In a separate measurement with finer frequency resolution, we resolved these modes at $422$ kHz and $621$ kHz, with estimated mean phonon excitation $\bar{n}=1.7(4)$ and $1.8(2)$ respectively. Their higher residual excitations can be attributed to enhanced sensitivity to electric-field noise from the trap electrodes~\cite{Kalincev2021,Boldin2018}, distinguishing them from adjacent modes in the sideband spectrum.  More details for the above results can be found in Supplementary Information~\cite{sm}. In comparison, for the Doppler-limited temperature $T_D \approx 0.5$~mK, the expected mean phonon excitation is $\bar{n}_{D}\approx\frac{k_{\mathrm{B}}T_D}{\hbar\omega_m}$, ranging from approximately 5 to 350 quanta of excitation, depending on the mode frequencies, as shown in Fig.~\ref{fig2}b. Such a substantial thermal excitation is qualitatively consistent with the diminished sideband asymmetry in Fig.~\ref{fig2}a. Thus, the vast majority of the collective motional spectrum is cooled below a single excitation, and the measured macroscopic modes are cooled well below the Doppler-limited values.

\begin{figure*}[tb] 
    \centering
    \includegraphics[width=2.0\columnwidth]{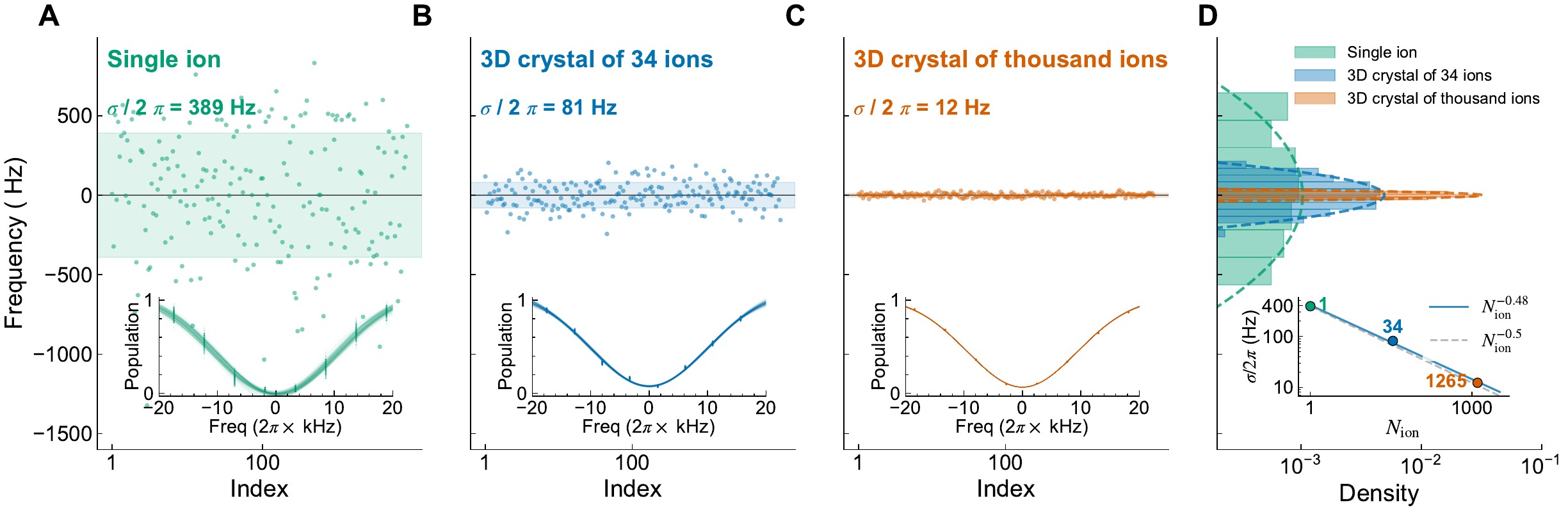}
    \caption{\textbf{Ensemble-enhanced precision in transition spectroscopy.} Probability distributions of the extracted optical-transition frequency become progressively narrower with increasing ion number for a single ion (green) in \textbf{a}, a 34-ion crystal (blue) in \textbf{b}, and a 3D Coulomb crystal containing 1,265 ions (orange) in \textbf{c}, measured under identical optical driving conditions. All fitting results are shown in the inset figure.  The frequency deviations are referenced to the mean value $\langle f\rangle$ for each dataset, and the corresponding standard deviations are $\sigma / 2 \pi =389  ~\mathrm{Hz}$, $81  ~\mathrm{Hz}$, and $12 ~\mathrm{Hz}$, respectively, with histogram shown in \textbf{d}. The inset of \textbf{d} shows the measured frequency uncertainty $\sigma$ as a function of ion number. The blue line indicates the fitting result $ N_{\mathrm{ion}}^{-0.48}$ scaling, and the dashed line indicates the expected standard quantum limit $ N_{\mathrm{ion}}^{-0.5}$ scaling, highlighting the improvement in spectroscopic precision enabled by the large three-dimensional ion ensemble.
}
    \label{fig4}
\end{figure*}

\noindent
\textbf{Coherent optical control and metrology}
We next implement coherent optical control on the laser-cooled 3D crystal by resonantly driving the $\lvert S_{1/2},m=-1/2\rangle\leftrightarrow\lvert D_{5/2},m=-1/2\rangle$ transition after initializing the ions in the $\lvert S_{1/2},m=-1/2\rangle$ state. We observe Rabi oscillations detected through global fluorescence collected by a PMT (Fig. \ref{fig3}a). In contrast,  with only Doppler cooling, the oscillations exhibit strongly reduced contrast and rapid damping, consistent with inhomogeneous broadening of the Rabi frequencies caused by thermal excitation of the collective motional modes. 
The residual decay after EIT cooling is predominantly attributed to the spatial inhomogeneity of the drive-laser intensity across the crystal. We verify this effect by imaging the entire ion crystal at various drive durations using an electron-multiplying charge-coupled device (EMCCD) camera. We can isolate 11 ions from the others in the images and perform partial site-resolved detection~\cite{sm}, and we find local Rabi oscillations remain in high contrast, with several examples shown in the same figure. By fitting laser beam parameters to the Rabi frequencies of the site-resolved ions,  we construct the Rabi-rate distribution across the entire crystal (Fig. \ref{fig3}c). Based on this model, we numerically simulate the ensemble Rabi oscillation with global detection (shown in Fig. \ref{fig3}a), well matched with our measurement data. Such spatial inhomogeneity could be substantially reduced using a flat-top beam profile with improved intensity uniformity across the crystal \cite{wang2025}. Despite the inhomogeneity, we can perform coherent optical control via a pulse sequence with varied durations and phases. As an example, we measure the ensemble optical coherence using Ramsey interferometry ~\cite{Degen2017,Wang2021}. We observe the fitted coherence time of $T_2=0.70(1)$~ms, limited primarily by a spatial magnetic-field gradient, which can be extended to $21.6(7)$ ~ms with dynamical decoupling sequences (Supplementary Information~\cite{sm}).

We next quantify the statistical precision of spin-resonant frequency measurement with varied ion numbers. Similar to the above Rabi oscillation experiments, here we instead scan the laser frequency at the duration of maximum contrast.  Spectroscopy is performed on a single ion and 3D crystals with 34 and  1,265 ions, using a step-size of 5 kHz across $\pm 20$ kHz. For each dataset, 50 repetitions are randomly selected from 200 original trials, and the corresponding spin-resonance frequency is extracted from curve fitting. The statistics of the measured frequencies from 200 resampling datasets give standard deviations of $\sigma/2\pi=389$ Hz, $81$ Hz, and $12$ Hz for the single ion, 34-ion crystal, and 1,265-ion crystal, respectively, as shown in Fig.~\ref{fig4}. Across varied ion numbers, we use a Rabi frequency of approximately 15 kHz and an interaction time of $34~\mu\mathrm{s}$, consistent within measurement uncertainties. The result shows a 32-fold reduction in statistical uncertainty for the 1,265-ion crystal compared with the single ion measurements, and demonstrates the enhanced spectroscopic precision enabled by the large three-dimensional ensemble, which scales as $N_{\mathrm{ion}}^{-0.48}$ (Fig.~\ref{fig4}d), nearly reaching standard quantum limit $N_{\mathrm{ion}}^{-0.5}$.

\noindent \textbf{\large{}Discussion}{\large\par}
\noindent 

We have cooled a three-dimensional Coulomb crystal of 1,265 ions to near-ground-state occupation across a continuous vibrational spectrum, and demonstrated global optical initialization, coherent control and collective readout. A coherence time of 21.6~ms under dynamical decoupling and a 32-fold reduction in the statistical uncertainty of the carrier-transition frequency compared with a single ion have been achieved in the 3D ion crystal. Our work validates that the dense mode spectrum of the 3D ion crystal is no longer an obstacle to coherent optical control~\cite{Thompson2015, Morigi2025}. The techniques demonstrated in this work can be easily extended to even larger crystals, as only the mode density increases in the vibrational spectrum, while the cross-section of the crystal and the corresponding optical power for global control grow as	$N_{\rm ion}^{2/3}$. The crystal self-assembles with no vacancies and requires no rearrangement step~\cite{Hornekaer2002, Drewsen1998}, persists for hours in the cryogenic environment~\cite{Spivey2022}, and has a mean inter-ion spacing comparable to that of neutral-atom tweezer arrays and therefore compatible with the single-site addressing techniques developed for those platforms~\cite{Bluvstein2024,Wang2016Single,Barredo2018,Ma2026}.

These results open a path for 3D ion-crystal quantum technologies with optical control. For optical clocks, the large ion number directly enhances signal-to-noise ratio~\cite{Ludlow2015, Bollinger2015, Leibrandt2024, Keller2019, Akerman2025}, and the measured coherence time of 21.6~ms under dynamical decoupling approaches the interrogation timescales explored in typical trapped-ion optical clock experiments, suggesting that ensemble-based clock references are within reach. Recent advances shed light on improved optical clocks based on entangled ions~\cite{Eckner2023, Franke2023, Dietze2026}. In such a large ion ensemble, spin-squeezing could further enhance the clock stability.
For quantum simulation, the volumetric geometry naturally hosts 3D spin models including frustrated magnetism~\cite{kim2010, islam2013, Balents2010} and many-body localization~\cite{Chertkov2021}. The motional-mode structure can also be exploited to simulate spin-boson models with structured vibrational environments~\cite{Lemmer2018, sun_quantum_2025}. The demonstrated combination of broadband cooling, global coherent control, and dynamical decoupling provides the essential control toolbox for these applications. Future directions would include exploiting the collective enhancement in cooling large crystal~\cite{vybornyi2026}, region-resolved or even site-resolved imaging techniques in 3D crystals to enable correlation measurements in many-body dynamics, and extending volumetric imaging~\cite{Zhou2024} together with more precise ion number counting for example with machine learning techniques~\cite{Yin2025}. The structural degrees of freedom of large Coulomb crystals could also enable controlled studies of nonequilibrium dynamics and topological defects \cite{Ruffert2026}.

\begin{acknowledgments}
\emph{Acknowledgments.---}
We thank L.-M. Duan, Y.-K. Wu, L. You, and W.-g. Wang for helpful discussions. We thank Z.Y. Liu and CIQTEK for support with the apparatus. This work is funded by the Quantum Science and Technology-National Science and Technology Major Project (Grant No.~2021ZD0301603), the National Natural Science Foundation of China (Grant No. 92565306, No. 125B2098, No. 12604783), the Chinese Academy of Sciences (Grant No.~XDB1300000), the National Key Research and Development Program of China (Grant No. 2025YFE0217900), the Postdoctoral Fellowship Program of CPSF (Grant No. GZC20261828) and Anhui Province Science and Technology Innovation Tackle Plan Project.

\end{acknowledgments}

\bibliography{bibfile}

\newpage
\clearpage

\onecolumngrid      

\renewcommand\thefigure{S\arabic{figure}}
\setcounter{figure}{0}

\noindent

\begin{center}
\Large\bfseries
Supplementary Information for:
``Coherent optical control of a thousand qubits in a three-dimensional trapped-ion crystal''
\end{center}

This Supplementary Information provides additional experimental details and characterizations for the cooling and coherent-control results presented in the main text. We first present the molecular-dynamics simulation of the crystal and the robustness of the spectral quantities used in the main text against variations of the equilibrium configuration. We then describe the complete experimental sequence and the implementation and modeling of parallel-EIT cooling. We further characterize the spatial distribution of the coherent drive Rabi frequency across the three-dimensional crystal and show that this inhomogeneous optical coupling quantitatively accounts for the residual decay of the globally measured Rabi oscillation in Fig.~3a of the main text. In addition, we examine motional heating, spatial distribution of the spin-resonance frequency, and optical-qubit decoherence, indicating that these mechanisms would not limit the short-time coherent population dynamics while quantifying the motional and coherence properties relevant for millisecond-scale interrogation of the large ion ensemble.

\section*{I. Molecular-dynamics simulation of the crystal and its vibrational spectrum}

\begin{figure*}
    \centering
    \includegraphics[width=0.9\linewidth]{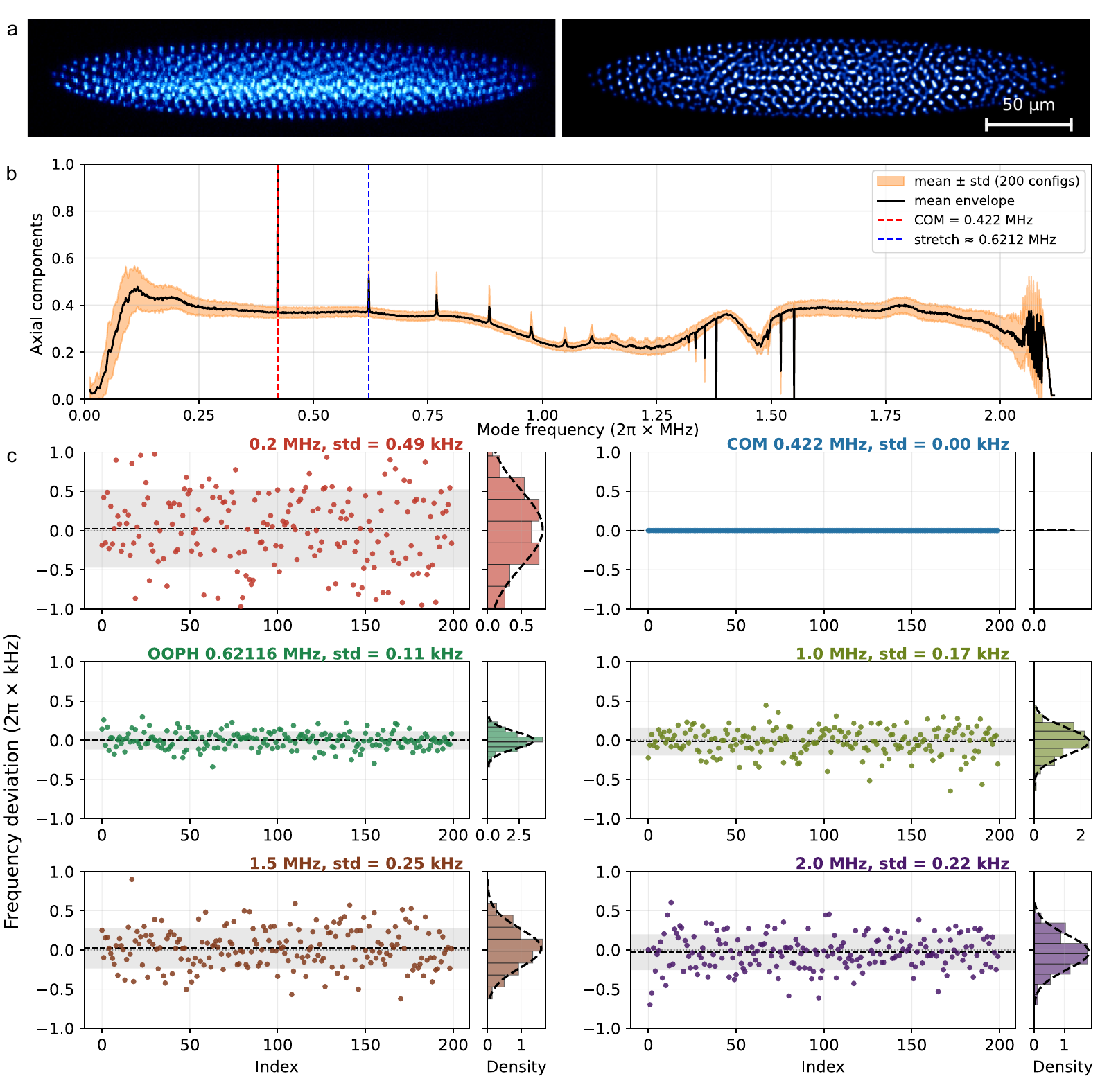}

    \caption{\textbf{Molecular-dynamics simulations of the 3D Coulomb crystal.} \textbf{a}, Fluorescence image of the three-dimensional Coulomb crystal used in the main experiment and a representative molecular-dynamics equilibrium configuration obtained using the experimentally measured trapping parameters, with scale bar of $50~\mu\mathrm{m}$. \textbf{b}, Mean axial mode components versus mode frequency, averaged over 200 independently obtained equilibrium configurations; the shaded region denotes the standard deviation across configurations. Dashed lines mark the axial center-of-mass (COM) and out-of-phase (OOPH) modes. \textbf{c}, Configuration-to-configuration fluctuations of representative mode frequencies across the 200 equilibrium configurations. Despite the frequency shifts of individual modes, the spectrum remains sufficiently dense that the experimentally relevant drive is expected to overlap with at least one motional mode for nearly every configuration. The OOPH mode shows a particularly small frequency spread, highlighting the enhanced stability of this global mode against microscopic crystal rearrangements.}
    \label{fig:S_MD}
\end{figure*}

In the dense three-dimensional crystal, the ions overlap along the imaging direction, so the ion number cannot be counted directly. Neither the equilibrium configuration nor the vibrational spectrum can be derived analytically. We therefore obtain the ion number, the configuration, and the spectrum from molecular-dynamics simulations in the harmonic trapping potential defined by the measured secular frequencies. The simulated equilibrium configuration reproduces the crystal shape and axial extent measured in the fluorescence image (Fig.~\ref{fig:S_MD}a).  This agreement fixes the ion number at $N_{\mathrm{ion}}=1{,}265\pm35$ and provides the configuration used to compute the mode spectrum of main-text Fig. 1c.

The calculated spectrum, however, depends on which metastable equilibrium configuration the crystal occupies, and this configuration is not controlled in the experiment. Treating the spectrum as a dense, quasi-continuous manifold is therefore justified only if the quantities on which that description relies remain representative across configurations. These quantities are the spectral envelope, the mode density that supports the 10-kHz sampling grid of the swept sideband spectroscopy, and the frequencies of the macroscopic axial modes. We verify their robustness by repeating the simulation for 200 independently initialized equilibrium configurations in Fig.~\ref{fig:S_MD}b. Across these configurations, the axial-projection envelope has a standard deviation of up to 17\% and of 3.4\% on average. Because the spectrum contains $1.8(6)$ modes per 1~kHz, spectral content remains within 1~kHz of every sampling point of the 10-kHz grid used in the main text, regardless of the configuration. Individual modes shift between realizations, whereas the axial centre-of-mass (COM) and out-of-phase (OOPH) modes show negligible frequency deviations. The frequency-grid sampling of the main text and the mode-resolved measurements described below are therefore unaffected by which configuration the crystal occupies.

\section*{II. Experimental sequence }

\begin{figure*}[h]
    \centering
    \includegraphics[width=0.6\linewidth]{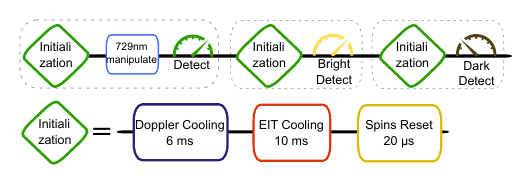}
    \caption{\textbf{Experimental sequence for motional cooling and quantum control.} The ions are first Doppler cooled, followed by parallel-EIT cooling and spin-state initialization. Subsequent 729 nm manipulation and fluorescence detection cycles are used for motional spectroscopy and coherent control. Bright- and dark-state references are interleaved to compensate detection fluctuations.}
    \label{figE:sequence}
\end{figure*}

The complete experimental sequence, illustrated in Fig.~\ref{figE:sequence}, consists of initialization, coherent manipulation, and fluorescence detection. The coherent manipulation implements the measurements described in the main text, including Rabi oscillations, motional spectroscopy, and coherence measurements. To suppress the influence of slow intensity drifts in the detection laser, experimental measurements are interleaved with reference measurements in which all ions are prepared in the bright and dark states, and the fluorescence signal is normalized accordingly.


The initialization comprises 6~ms of Doppler cooling, 10~ms of parallel-EIT cooling, and 20~$\mu$s of optical pumping, preparing the ions  to the $\ket{S_{1/2},m=-1/2}$ spin state.
Doppler cooling is performed using a $\pi$-polarized 397~nm beam together with the 866~nm repump laser. After parallel-EIT cooling as detailed below, the ions are optically pumped into the $\ket{S_{1/2},m=-1/2}$ state using a resonant $\sigma^{-}$-polarized 397~nm beam together with the 866~nm repump laser. 
\begin{figure*}[h]
    \centering
    \includegraphics[width=0.7\linewidth]{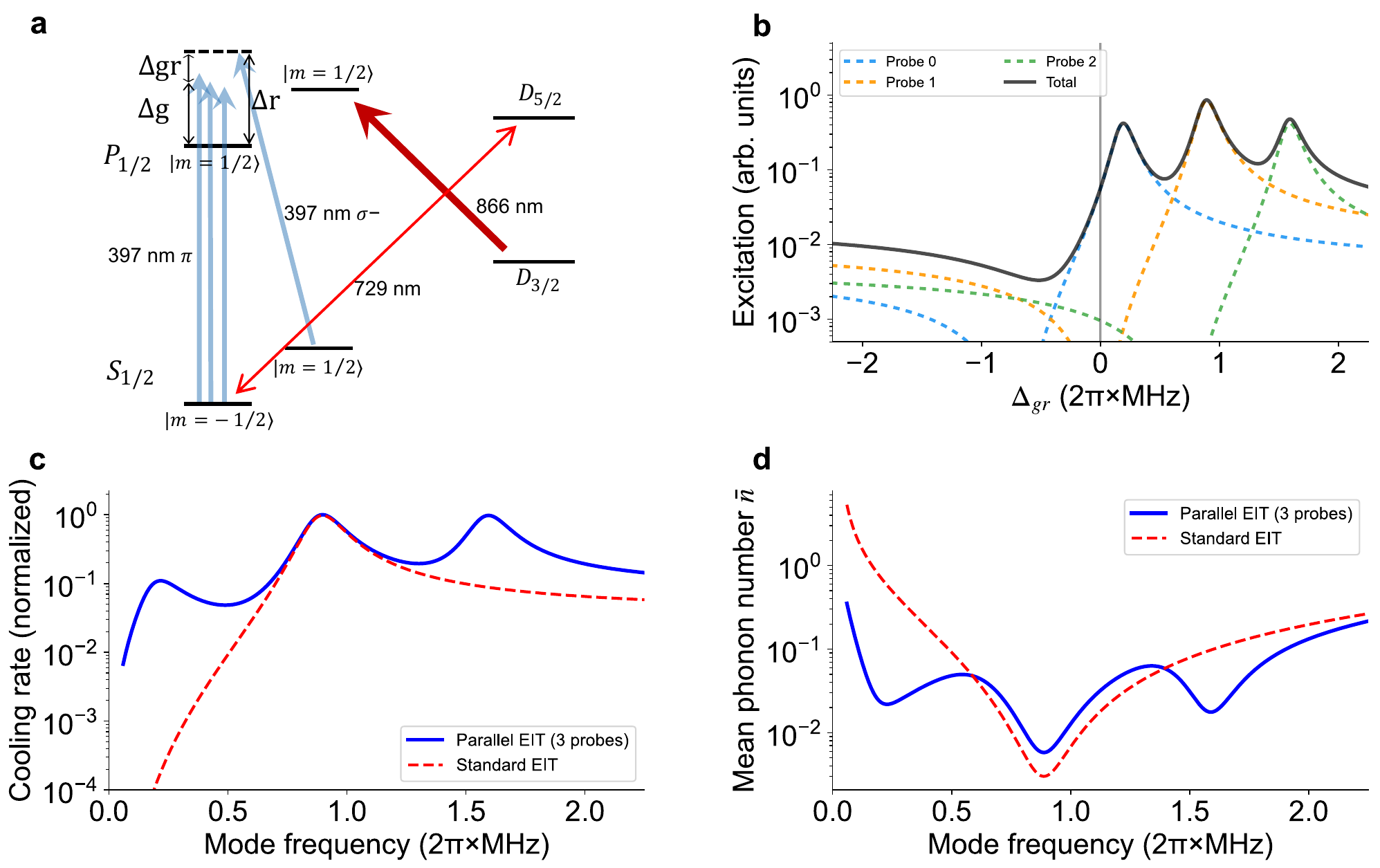}
    \caption{ \textbf{Model of parallel-EIT cooling.}
    \textbf{a,} Level scheme used for parallel-EIT cooling in $^{40}$Ca$^{+}$. 
    A $\sigma^{-}$-polarized 397~nm beam couples the $\ket{S_{1/2},m=-1/2}\leftrightarrow\ket{P_{1/2},m=-1/2}$ transition, while a multi-frequency $\pi$-polarized 397~nm beam addresses the $\ket{S_{1/2},m=+1/2}\leftrightarrow\ket{P_{1/2},m=-1/2}$ transition. 
    An 866~nm laser repumps population from the $D_{3/2}$ manifold.
    \textbf{b,} Calculated excitation spectrum for the experimentally relevant parameters. The three probe frequencies generate distinct Fano-like resonances whose superposition produces multiple transparency windows distributed across the vibrational spectrum. \textbf{c,} Calculated cooling rate as a function of motional frequency. The parallel-EIT configuration (blue) provides efficient cooling over a broad frequency range, whereas conventional single-resonance EIT cooling (red dashed curve) is effective only within a narrow spectral region near the AC Stark shift. \textbf{d,} Calculated steady-state mean phonon excitation as a function of motional frequency. The parallel-EIT configuration maintains low phonon excitations over a substantially larger fraction of the vibrational spectrum than conventional EIT cooling. The calculations use experimentally relevant laser parameters and illustrate the broadband-cooling mechanism enabled by multiple resonances.
}
    \label{figE:level}
\end{figure*}

For parallel-EIT cooling, a $\sigma^{-}$-polarized 397~nm beam serves as the coupling field and a multi-frequency $\pi$-polarized 397~nm beam provides the probe fields, as shown in Fig.~\ref{figE:level}a, forming a $\Lambda-$configuration. Both beams are detuned by approximately $\Delta_r=2\pi\times100~\mathrm{MHz}$ from the $S_{1/2}\rightarrow P_{1/2}$ transition resonance. The coupling field has a Rabi frequency of approximately $2\pi\times19~\mathrm{MHz}$, corresponding to an AC Stark shift of $\delta_{\rm AC}\approx2\pi\times0.9~\mathrm{MHz}$. The probe field contains three frequency components separated by approximately $2\pi\times0.7~\mathrm{MHz}$, each with a Rabi frequency of approximately $2\pi\times2~\mathrm{MHz}$.  Their frequencies are chosen following Ref.~\cite{Zhang2022} such that the resulting dark resonances are distributed across the motional spectrum (Fig.~\ref{figE:level}b). Both EIT beams have elliptical intensity profiles with $1/e^2$ diameters of approximately $200~\mu\mathrm{m}$ and $2~\mathrm{mm}$ along their short and long axes, respectively.
The cooling rate and steady-state mean phonon occupation of parallel-EIT are simulated in Fig.~\ref{figE:level} c\&d. Although the peak cooling rates are comparable, the steady-state occupations differ because probe components away from a given dark resonance introduce residual off-resonant scattering and heating. Parallel-EIT therefore trades a slightly higher minimum phonon occupation for a substantially broader cooling bandwidth.



\begin{figure*}[h!]
    \centering
    \includegraphics[width=0.6\linewidth]{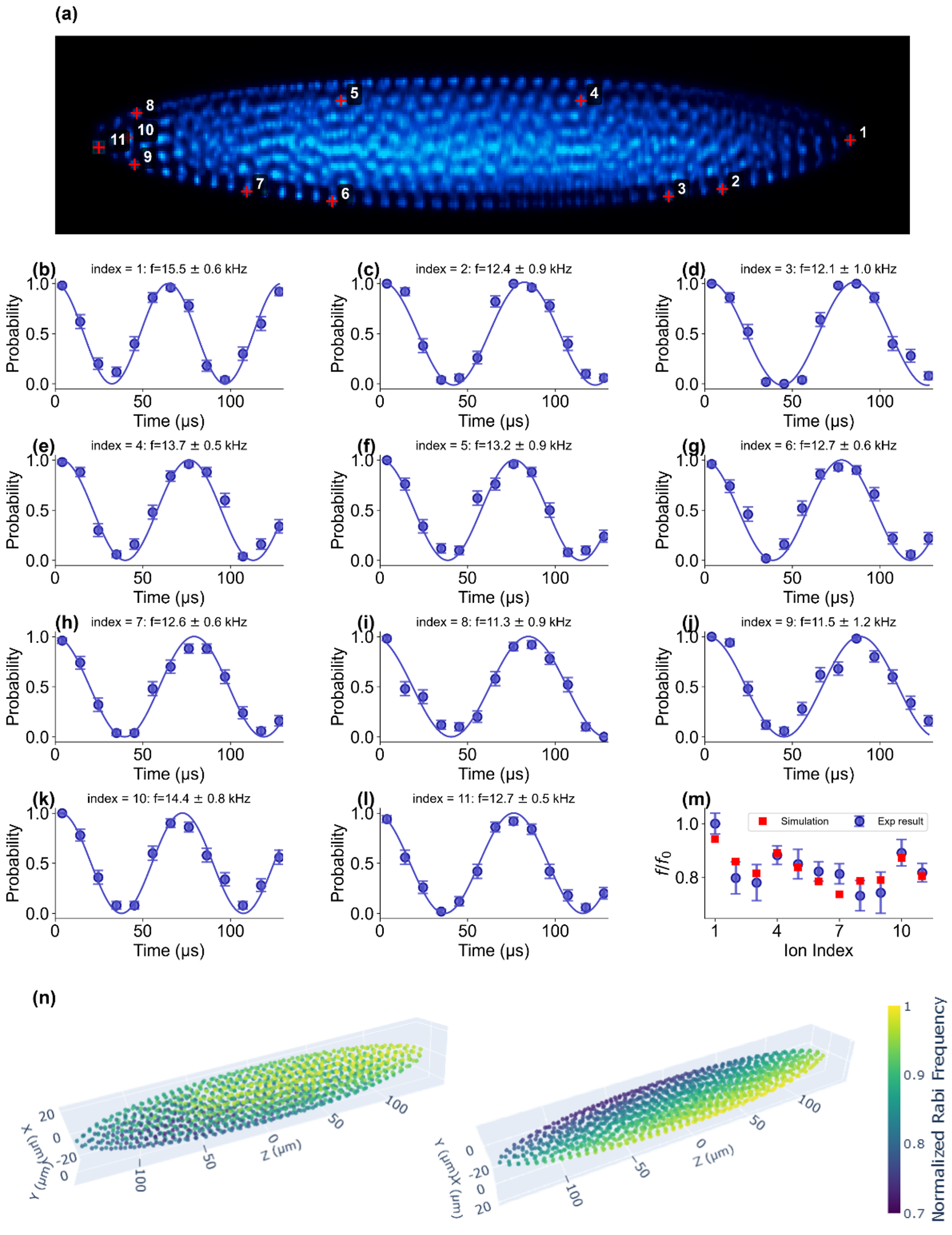}
    \caption{\textbf{Characterization and modeling of the spatial distribution of Rabi frequencies in the three-dimensional ion crystal.} Local Rabi oscillations measured at 11 spatially resolved ions within the crystal. The extracted Rabi frequencies are compared with a three-dimensional crystal model. The 729~nm driving beam is described by a Gaussian profile with a beam waist of $90~\mu$m. The beam waist and incidence direction are optimized by maximum likelihood fitting to the measured Rabi frequencies. The simulated Rabi frequencies agree with the experimental results within the measurement uncertainties, as shown in panel \textbf{(m)}. The lower panels show the calculated three-dimensional distribution of the normalized Rabi frequency for all ions in the crystal under Gaussian beam illumination. 
}
    \label{fig:S_rabi}
\end{figure*}

\section*{III. SPATIAL DISTRIBUTION OF THE COHERENT DRIVE RABI FREQUENCIES}



We characterize the beam waist and propagation direction by maximum-likelihood fitting to the measured Rabi frequencies of the 11 resolved ions (shown in Fig.~\ref{fig:S_rabi}a) together with the measured crystal geometry, yielding a beam waist of approximately 90~$\mu$m and beam propagation direction (Fig.~\ref{fig:S_rabi}). The resulting model reproduces the measured local Rabi frequencies within their experimental uncertainties and provides the Rabi frequency of every ion in the crystal.
The model yields a three-dimensional distribution of the normalized Rabi frequency across the entire crystal (Fig.~\ref{fig:S_rabi} (m))). 
Using this distribution, we further simulate the globally detected carrier dynamics by averaging the coherent evolution over the entire ensemble.  The simulation quantitatively reproduces the measured ensemble decay of the main text Fig.~3a.
To suppress the spatial distribution of the coherent drive Rabi frequencies, a more homogeneous optical profile, for example a flat-top beam, will be beneficial~\cite{wang2025}.

\section*{IV. MOTIONAL HEATING AND ITS EFFECT ON COHERENT CONTROL}

\begin{figure*}[h]
    \centering
    \includegraphics[width=1.0\linewidth]{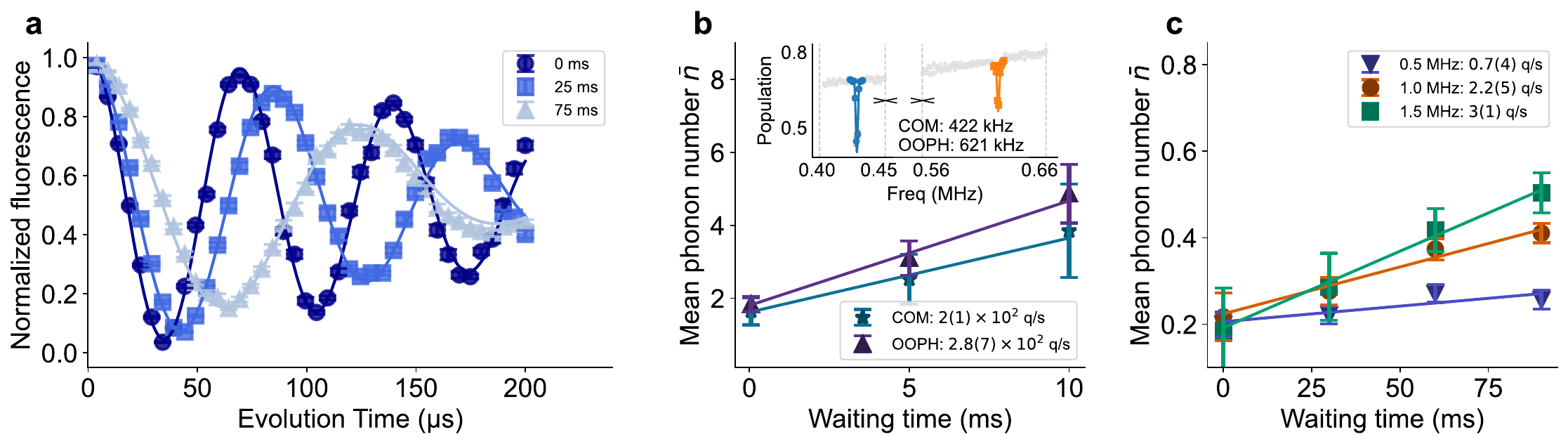}
    \caption{\textbf{Motional heating after parallel-EIT cooling.} \textbf{a}, Representative spin-resonance population dynamics measured after waiting times of 0, 25, and 75~ms following parallel-EIT cooling. \textbf{b}, Mean phonon occupation of the axial centre-of-mass (COM) and out-of-phase (OOPH) modes as a function of waiting time. The inset shows the sideband spectrum resolving the two modes at approximately 422 kHz and 621 kHz. \textbf{c}, Mean phonon occupation measured at representative mode frequencies of \{0.5, 1.0, 1.5\}~MHz.
 }
    \label{fig:S_EIT}
\end{figure*}

We next characterize the motional state following parallel-EIT cooling and assess whether heating contributes to the residual decay of the global Rabi oscillation. Following the cooling sequence, the ions are held in the $\ket{S_{1/2}, m=-1/2}$ state for a variable waiting time before  the coherent Rabi drive. Figure~\ref{fig:S_EIT}a shows representative Rabi oscillations after waiting times of 0, 25, and 75 ms. 
Even after 75 ms, the oscillation contrast remains above 80\%, demonstrating that the low-excitation motional state persists well beyond the timescale relevant for coherent manipulation.


To quantify such a heating effect, we measure the heating rates of representative macroscopic and local modes. The two axial macroscopic modes, the centre-of-mass (COM) mode and out-of-phase (OOPH) modes, are resolved in a higher-resolution frequency scan around the low-frequency motional spectrum at approximately $2\pi\times422$~kHz and $2\pi\times621$~kHz, respectively, as shown in the inset of Fig.~\ref{fig:S_EIT}b. We then probe the motional excitation of this two modes via red- and blue-sideband evolution after variable waiting times. The extracted mean phonon occupations increase approximately linearly with waiting time, and linear fits give heating rates of $2(1)\times10^{2}~\mathrm{quanta/s}$ for the COM mode and $2.8(7)\times10^{2}~\mathrm{quanta/s}$ for the OOPH mode.

We additionally measure the heating at representative frequencies 0.5, 1.0, and 1.5~MHz, obtaining corresponding heating rates of $0.7(4)$, $2.2(5)$, and $3(1)~\mathrm{quanta/s}$, respectively (Fig.~\ref{fig:S_EIT}c). These rates are substantially lower than those of the macroscopic modes. 

\section*{V. Spatial distribution of the spin-resonance frequency}
\begin{figure*} 
    \centering
    \includegraphics[width=1.0\linewidth]{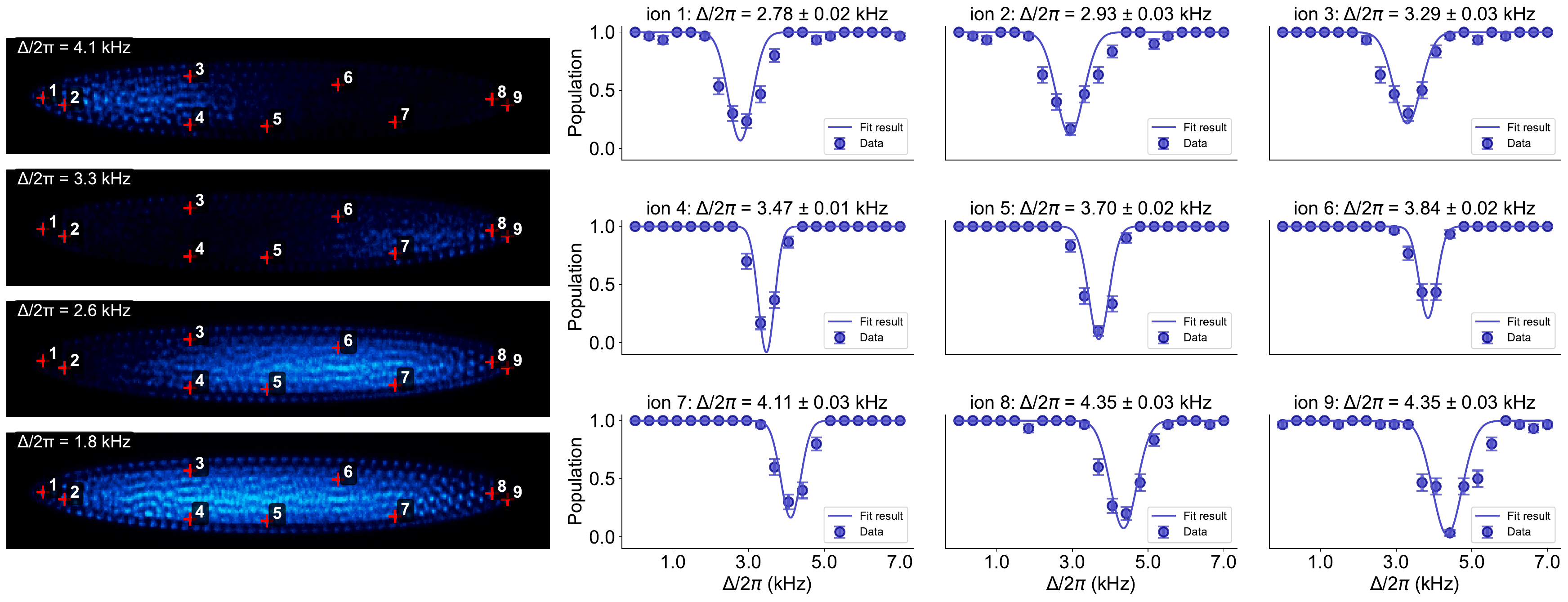}
    \caption{ \textbf{Characterization of the spatial distribution of spin-resonance frequencies in the three-dimensional ion crystal.} Spin-resonance spectra measured at different spatial positions within the crystal. The local transition frequencies are extracted from Gaussian fits of spatially resolved ions with stable positions during the measurement. The frequency axis denotes relative frequency variation of the single-pass acousto-optic modulator drive frequency, which directly determines the frequency detuning of the 729 nm laser. The observed position-dependent frequency shifts $\Delta/2\pi$, amounting to 1.57 kHz over the 281 $\mu$m crystal length, reveals a spatial distribution of the spin-resonance frequency across the crystal. }
    \label{fig:gradient}
\end{figure*}
A spatial variation of the spin-resonance frequency across the crystal can broaden the optical transition frequency and thus constitute another potential mechanism for the residual decay of the global Rabi oscillation. To examine this effect, we select several spatially resolved ions that remain positionally stable during the measurement (Fig.~\ref{fig:gradient}) and record their spin-resonance spectra, with a reduced Rabi frequency of $\Omega/2\pi\approx0.7$~kHz to narrow the power broadening of the spectra and thereby improve the frequency resolution.  Each spectrum is fitted with a Gaussian profile to extract the local spin-resonance frequency. The measured transition frequencies exhibit a clear spatial dependence.

The resonance frequency difference between the two edges of the crystal is $\Delta f = 1.57~\mathrm{kHz}$ over an axial distance of $L = 281~\mu\mathrm{m}$. Using the Zeeman sensitivity of the $|S_{1/2},m=-1/2\rangle \rightarrow |D_{5/2},m=-1/2\rangle$ transition, $\frac{\partial f}{\partial B}=(g_Dm_D-g_Sm_S)\frac{\mu_B}{h}$, the frequency difference corresponds to a magnetic-field variation of approximately $\Delta B \approx 2.9~\mathrm{mG}$ across the crystal. This frequency spread is much smaller than the typical resonant Rabi frequency used for global coherent manipulation ($\Omega/2\pi\approx15$~kHz). Therefore, it does not contribute significantly to the residual decay of the global Rabi oscillation in Fig.~3a in the main text.

\section*{VI. Characterization of spin coherence}

We examine whether optical-qubit decoherence limits the short-time coherent population dynamics of the Rabi-oscillation measurements.
Coherence measurements are performed on the optical qubit formed by the $4S_{1/2}(m_J=-1/2)$ and $3D_{5/2}(m_J=-1/2)$ states of $^{40}$Ca$^{+}$. Following state preparation, a resonant $\pi/2$ pulse on the 729-nm quadrupole transition prepares the superposition state $(|S_{1/2}\rangle+|D_{5/2}\rangle)/\sqrt{2}$. After an evolution time $t$, a second $\pi/2$ pulse is applied and the final population is measured by global PMT fluorescence detection. For spin-echo measurements, a single resonant $\pi$ pulse is applied at $t/2$. Higher-order dynamical decoupling uses XY4 and XY8 sequences \cite{Wang2016} with equally spaced $\pi$ pulses ($X-Y-X-Y$ for XY4 and $X-Y-X-Y-Y-X-Y-X$ for XY8). The coherence is extracted from the decay curve and fitted with a Gaussian function, $C(t)=A\exp[-(t/T_2)^2]+B$, where $T_2$ denotes the coherence time. The same measurements are performed on a single ion under identical experimental conditions, including the optical transition, laser system, magnetic-field environment, and pulse sequences. 

\begin{figure*}
    \centering
    \includegraphics[width=0.8\linewidth]{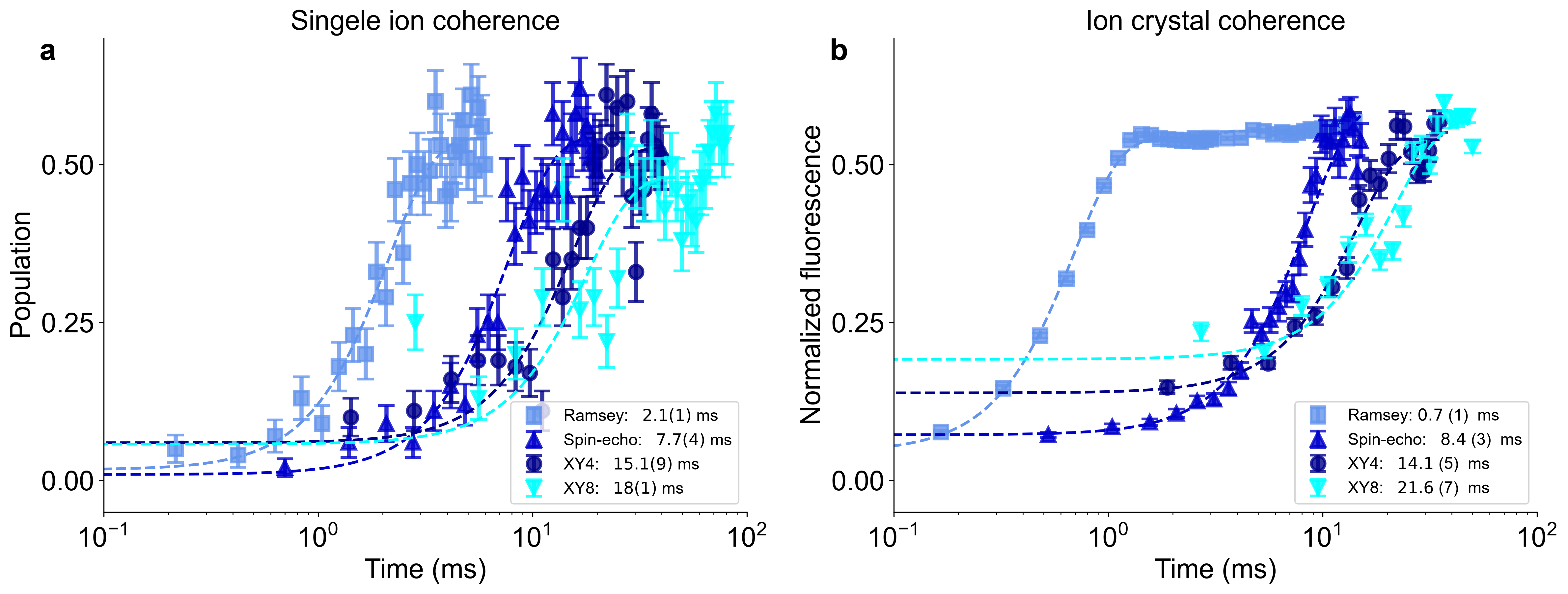}
    \caption{  \textbf{Comparison of coherence times between a single ion and a three-dimensional ion crystal.} \textbf{a,}  Coherence time measurements of a single $^{40}$Ca$^+$ ion under Ramsey, spin-echo, XY4, and XY8 dynamical decoupling sequences. The extracted coherence times are $T_2=2.1(1)$~ms, $7.7(4)$~ms, $15.1(9)$~ms, and $18(1)$~ms, respectively. \textbf{b,} The corresponding coherence decay measurements for the three-dimensional ion crystal, with extracted coherence times of $T_2=0.70(1)$~ms, $8.4(3)$~ms, $14.1(5)$~ms, and $21.6(7)$~ms. The comparable refocused coherence times demonstrate that scaling from a single ion to a large ion crystal does not additionally introduce notable decoherence.}
    \label{fig:S_coherence}
\end{figure*}

Figure~\ref{fig:S_coherence} shows the coherence measurements for the three-dimensional ion crystal. The extracted coherence times extend from $T_{2}^{\mathrm{Ramsey}} = 0.70(1)\,\mathrm{ms}$ to $T_{2}^{\mathrm{echo}} = 8.4(3)\,\mathrm{ms}$, $T_{2}^{\mathrm{XY4}} = 14.1(5)\,\mathrm{ms}$, $T_{2}^{\mathrm{XY8}} = 21.6(7)\,\mathrm{ms}$. For a single ion measured under the same conditions, the corresponding values are $T_{2}^{\mathrm{Ramsey}} = 2.1(1)\,\mathrm{ms}$, $T_{2}^{\mathrm{echo}} = 7.7(4)\,\mathrm{ms}$, $T_{2}^{\mathrm{XY4}} = 15.1(9)\,\mathrm{ms}$ and $T_{2}^{\mathrm{XY8}} = 18(1)\,\mathrm{ms}$. The refocused coherence of the crystal is thus comparable to that of a single ion, showing that scaling to the three-dimensional ion crystal does not introduce substantial irreversible decoherence. The additional dephasing in the extended system is predominantly static or low-frequency and can be efficiently suppressed by dynamical decoupling, extending the crystal coherence time to the interrogation timescales relevant for an ensemble-based clock reference~\cite{Akerman2025}. 

\end{document}